\documentclass[journal,comsoc]{IEEEtran}
\usepackage[nointegrals]{wasysym}
\usepackage[T1]{fontenc}
\usepackage{graphicx}
\usepackage{booktabs}
\usepackage{rotating}
\usepackage[flushleft]{threeparttable}
\usepackage[utf8]{inputenc}
\usepackage[english]{babel} 
\usepackage{amssymb}
\usepackage{amsmath}
\usepackage{txfonts}
\usepackage{mathdots}
\usepackage{mdframed}
\usepackage{mwe}
\usepackage{multirow}
\usepackage{listings}
\usepackage[justification=centering]{caption}

\lstdefinestyle{mystyle}{
    basicstyle=\footnotesize,
    breakatwhitespace=false,
    breaklines=true,
    captionpos=b,
    keepspaces=true,
    numbers=left,
    numbersep=5pt,
    showspaces=false,
    showstringspaces=false,
    showtabs=false,
    tabsize=2
}
\begin{document}
\title{Cerberus: A Blockchain-Based Accreditation and Degree Verification System}

\author{\IEEEauthorblockN{Aamna Tariq\IEEEauthorrefmark{1},
Hina Binte Haq \IEEEauthorrefmark{2} , Syed Taha Ali\IEEEauthorrefmark{3}} \\
\IEEEauthorblockA{ School of Electrical Engineering and Computer Sciences (SEECS), National University of Sciences and Technology (NUST), Islamabad, Pakistan.\\ \IEEEauthorrefmark{1} \IEEEauthorrefmark{2} contributed equally to the paper.\\
\IEEEauthorrefmark{1}atariq.msit16seecs@seecs.edu.pk,
\IEEEauthorrefmark{2}hhaq.dphd18seecs@seecs.edu.pk,
\IEEEauthorrefmark{3}taha.ali@seecs.edu.pk}}

\maketitle
\begin{abstract}
Credential fraud is a widespread practice that undermines investment and confidence in higher education systems and bears significant economic and social costs. Legacy credential verification systems are typically time-consuming, costly, and bureaucratic, and struggle against certain classes of credential fraud. In this paper, we propose a comprehensive blockchain-based credential verification solution, Cerberus, which is considerably more efficient, easy and intuitive to use, and effectively mitigates widespread manifestations of credential fraud. Cerberus also improves significantly upon other blockchain-based solutions in the research literature: it adheres closely to the existing credential verification ecosystem, it addresses a threat model informed by real-world fraud scenarios. Moreover, Cerberus uses on-chain smart contracts for credential revocation, and it does not entail students or employers to manage digital identities or cryptographic credentials to use the system. We prototype our solution and describe our attempt to design an online verification service with a rich feature set, including data privacy, transcript verification, and selective disclosure of data. We hope this effort contributes positively to towards alleviating the problem of fake
credentials.
\newline
\newline

\centerline{\textbf{Managerial Relevance Statement}}

There is widespread recognition that the blockchain can effectively combat the problem of fake credentials and significantly improve the credential verification process. There are several efforts underway in this domain, both in industry and ideas proposed in the research literature. However, to the best of our knowledge, what is most lacking is a rigorous effort to tailor these solutions to existing practices of credential fraud, and to mitigate usability issues on the part of users.

Our system will enable practitioners (universities, accreditation bodies, employers) to undertake the following: 1) to concretely visualize the workings and benefits of such a system and work towards integrating it into existing credential verification ecosystems, and 2) to adapt and customize the rich feature set of our solution to cater to their own requirements.

\end{abstract}

\section{Introduction}

The value of education to society cannot be overstated: education plays a pivotal role in enabling social mobility \cite{socialmobility} \cite{socialmobility1}, it contributes to economic uplift \cite{ecogrowth} \cite{ecogrowth1} \cite{ecogrowth2}, and it promotes political stability and reform \cite{schweinhart} \cite{verba} \cite{reimers}. Education is also a critical factor for innovation to meet local and international challenges and opportunities \cite{humancapital}. For these reasons, education typically ranks near the top of government spending priorities in developed countries, and often absorbs over a fifth of total government spending in the public sector \cite{corredu}.

Corruption in the domain of education can have wide-ranging and detrimental effects. In 2013, in its global corruption report on education, Transparency International extensively documented various instances of such phenomena, particularly in developing countries where such fraud is pervasive and systemic. Corrupt practices range over a wide spectrum including bribery and nepotism in admissions and in examinations results, a culture of teacher absenteeism, deteriorating quality of education, misappropriation of funds, ghost schools existing only on paper, diploma mills which issue fake academic degrees, and compromised accreditation services \cite{ti}.

In this paper, we specifically focus on the problem of fake academic credentials. With increasing social pressure to outperform peers in highly competitive job markets, many applicants misrepresent their qualifications to make their curriculum vitae more appealing to employers. An academic credential typically requires considerable investment in terms of time, effort, and funding, and in turn it confers a certain prestige on the bearer and opens new opportunities to them. Fake degrees essentially enable third parties a `free ride' on these benefits \cite{gilles}.

Falsified credentials range from high school diplomas to doctorate degrees. There have been publicized instances where investigators have successfully procured fake degrees on behalf of pet cats \cite{cats} and dogs  \cite{dogs}. According to one study, over a third of potential candidates admit to falsifying or `enhancing' their qualifications for a job application \cite{lor}. A professor in South Africa was recently assassinated for exposing a syndicate producing fake PhD degrees \cite{unizulu}.  There is a marked scarcity of research on the scope of this problem, but a conservative estimate suggests that there are more than 5000 unrecognized universities and diploma mills operating worldwide, and issuing over 200,000 fake degrees annually with revenues in excess of \$1 billion \cite{ezell}.

These unethical practices not only discourage individual investment in education but also damage the value, credibility and reputation of a higher education system. Moreover, there may be considerable social harm: apart from a compromised sense of ethics, a fake degree holder will likely not possess the requisite expertise in his field (achieved through rigorous training and evaluation), thereby posing a real danger in certain domains. For instance, subjects implicated in fake degree scandals include doctors, nurses \cite{doctors} \cite{nurses} , pilots \cite{pilots} \cite{piapilots}, and even politicians and government ministers \cite{politicians} \cite{russia} \cite{jordan}.

The traditional defense against fake credentials is stringent verification procedures. In certain countries, this task is undertaken at a national level by government bodies, such as the Academic Degrees and Graduate Education Development Center in China \cite{chinaqual}, the Higher Education Commission in Pakistan \cite{hecver}, and the Higher Attestation Commission in Russia \cite{russiaattest}. In some territories, accreditation is overseen by non-governmental agencies, such as the Council for Higher Education Accreditation (CHEA) in the United States \cite{chea}, and Tertiary Education Quality and Standards Agency (TEQSA) \cite{teqsa}, Australia's independent national quality assurance and regulatory agency for higher education. There also exist wide-ranging international efforts such as the Hague Apostille Treaty of 1961, whereby citizens can have their credentials certified by a designated authority so that they may be recognized in 116 signatory countries \cite{hague}. Another facility is the IAU World Higher Education Database (WHED) which provides comprehensive information on accredited higher education systems and credentials around the world \cite{whed}.

However, most credential verification mechanisms are opaque, cumbersome, time consuming, and costly. According to a survey, one in three employers in the UK do not request candidates for their degree certificates; and, of those who do, 76\% of employers assume the certificates are legitimate and do not verify their authenticity \cite{independent}. Foregoing diligent background checks entails costs in terms of lost time, performance, and investment: in the US, the Department of Labour reports that forged credentials and doctored qualifications cost companies an average of \$40,000 per bad hire \cite{badhire}.

In recent years, the blockchain has been promoted as a promising new technology for transparency and data integrity in a variety of domains, including payment settlements \cite{ripple} \cite{stellar}, copyright protection \cite{binded} \cite{ascribe}, data notarization \cite{bnotary} \cite{proofofexistence}, digital government \cite{swedenland} \cite{smartdubai} \cite{e-Gov} \cite{e-Chaina}, health care \cite{medrec}, \cite{medshare}, \cite{e-health} \cite{clinical}, logistics and tracking \cite{blockfreight} \cite{everledger}, and secure elections \cite{agora} \cite{followmyvote}. Due to its distributed architecture and its reliance on cryptography, the blockchain offers strong guarantees on accountability, accessibility and data immutability, which is particularly suited to scenarios with multiple untrusted parties. Various efforts also advocate using the blockchain to combat the fake degree problem \cite{aversafe} \cite{bc4edu} \cite{EduCTX}, but as yet there are few rigorous efforts \cite{blockcerts} in this direction.

We attempt to address this deficiency. Our paper makes the following contributions:

\begin{enumerate}

\item We propose a comprehensive blockchain-based solution for easy and efficient verification of academic credentials. We describe a solution architecture which integrates seamlessly with typical credential management ecosystems and we devise a threat model informed by real-world fraud scenarios,

\item We prototype our solution and describe our attempt to design an online verification service with a rich feature set, including data privacy, transcript verification, and selective disclosure of data,

\item We propose a novel credential revocation mechanism. This is a distinct contribution in its own right. 
We suggest a simple and practical on-chain revocation mechanism which leverages smart contracts.

\end{enumerate}
We describe here briefly a representative scenario which motivates our solution: an Accreditation Authority operates and maintains a permissioned blockchain in partnership with universities and watchdog organizations. When a student, Alice, graduates, her university issues her a physical degree certificate and also add her details on the blockchain platform. This certificate also contains a QR code which allows Alice, or any other party, to verify her credentials in real-time from the blockchain using a smartphone app. This is similar to national visa verification services such as VEVO in Australia \cite{vevo}, the Employer Checking Service in the UK \cite{ecsuk}, and E-Verify in the US \cite{everifyus}, which allow employers to check the visa status of job applicants via a Web portal. Alice can even paste a QR code on her resume, thereby enabling prospective employers to verify her details independent of the certificate. In case her degree is revoked, the university will make corresponding entries in the blockchain which will be revealed when Alice's QR code is scanned.

This solution has notable differences with prior work: first, our effort is on engineering a solution that preserves the existing ecosystem and maintains key security properties such as guarantees on data privacy, integrity, and revocation. Second, it approximates and improves upon the real world process flow for credential management and verification. Furthermore, we focus on usability: our solution is relatively easy to use since verification can be done by scanning a QR code and neither Alice nor her employer are required to personally interact with the blockchain or maintain secret keys.

The rest of this paper is organized as follows: in Section II Background, we discuss the ground realities regarding academic credential fraud throughout the world and also discuss the higher education Ecosystem specific to Pakistan. In Section III we discuss the architecture and flow of the Proposed System. In Section IV we enumerate implementation details of our prototype of the system. In Section V Discussion, we highlight and explain how our system achieves the various security properties. In Section VI we compare and contrast our endeavour with similar systems and related work. In Section VII we delineate future direction for research and conclude our discussion in Section VII. 
\section{Background}

\subsection{The Varieties of Credential Fraud}
\label{sec:varietiesofcredentialfraud}

Credential fraud has been around since at least the fourteenth century \cite{ezell}, and there is considerable evidence that degrees were widely sold in German universities in the 18th century \cite{zaretsky}. However, this phenomenon gained rapid traction in the 20th century due to two main drivers: first, as Johnson convincingly argues, increasing global competition in job markets has given rise to a widespread culture of \emph{credentialism}, with employers `overly relying on degrees as proof of job competency', even for low-to-moderate skill positions. This practice likely contributes considerably to the black market for fake credentials \cite{credentialism}.

Second, the 20th Century marked the ascendance of the ``for-profit'' education model in schools and universities, whereby academic excellence and integrity had to contend increasingly with economic and business interests. This situation was further complicated by rapid expansion higher education institutions in the form of distance learning programmes, flexible and distributed learning modes, branch campuses, franchising, and credit transfer schemes \cite{transnationaledu}. The distributed and transnational nature of these schemes makes it considerably more difficult to enforce independent checks on quality and integrity.

The result is a pervasive and thriving culture of credential fraud and a billion dollar industry \cite{ti}. Whereas concrete figures on credential fraud are not available \cite{wenr}, some investigations reveal the alarming scope of this trend \cite{scourge}. For instance, in the US, home to the largest number of diploma mills in the world \cite{verifilereport}, Ezell et al. document that the number of fake PhD degrees purchased every year exceeds 50,000, outnumbering the 40,000-45,000 legitimate PhDs awarded by universities \cite{ezzelbook2005}. One diploma mill, operated by Americans with offices in Europe and the Middle East, has sold more than 450,000 degrees with revenues exceeding US \$450,000,000. \cite{ezzelbook2012}

In Europe, the UK is believed to host the largest number of diploma mills \cite{verifilereport}. A prominent example was the University of Wales, the second-largest university in the country with a 120-year history, which had 70,000 students enrolled in 130 colleges around the world. After multiple scams and administrative failures were uncovered, the registrar resigned and the university shuttered its highly profitable degree validation program, which accounted for nearly two thirds of institutional revenue \cite{univofwales} \cite{walesstats}.

Fraud is also rampant in the developing world. According to one estimate an alarming half of all high school transcripts in overseas university admissions applications by Chinese students are falsified \cite{wenr}. This problem is also very widespread in India \cite{thehindu} \cite{indiamedical}, and trafficking in fake certificates has been described as a `pan-India' crime \cite{rediffnews}. A 2015 study found that one in nine politicians in the lower house of the Russian parliament possessed a plagiarized or fake degree. In Indonesia, a task force was set up by the government in 2015 to crack down on fake degrees issued specifically to politicians \cite{politicians}.

Here we broadly classify various categories of credential fraud:

\paragraph{\textbf{Document Fraud}} typically involves illegal counterfeits, deceitful alteration of legitimate credentials (modification of name, signatures, degree, details, etc.), or complete fabrications (using fake logos, seals, and serial numbers) \cite{cvc}. This category also includes doctored or misleading translations and evaluations of credentials.

A recent example is the case of degree shops that have recently sprung up on the Syrian-Turkish border, where merchants exploit desperate Syrian migrants and refugees by selling them forged documents on their way to Europe. A high school diploma reportedly costs USD \$600, whereas a university degree can be as much as USD \$2,500 \cite{nytimes}.


\paragraph{\textbf{Institutional Fraud}} refers to the case where staff within institutions are compromised \cite{univofwales}. Such fraud may involve the university registrar or other officials  who create an illegitimate credential which is retroactively appended in the official record of the university. This tactic is more reliable than document fraud because the credential itself is authentic and can usually withstand cursory scrutiny because it is backed by university records.

A prominent example is the case of Busoga University in Uganda which was investigated in 2016 for issuing more than 1,000 ``premium-tuition'' degrees to South Sudanese students, most of them military officers seeking easy degrees to secure government positions \cite{uganda}.


\paragraph{\textbf{Diploma Mills}} sell fake credentials from fictitious universities and lead the mass market in credential fraud. These bodies operate in a highly structured and sophisticated manner, with a corporate culture including a dedicated marketing and sales teams, and offer customised "products" to buyers. These mills often maintain immaculate websites for fictitious universities \cite{axact}.

A recent example is the international scandal of Axact, widely considered the largest degree scam yet, where a Pakistan-based company that operated a web of more than 370 diploma mills which collectively earned millions of dollars in revenue by selling fake degrees and certificates of hundreds of fictitious universities to clients worldwide \cite{axact}. Axact have also been known to extort their customers for funds after making sales to them by threatening to reveal that their credentials were bogus \cite{axactextort}.



\paragraph{\textbf{Accreditation Fraud}} refers to the case where the accreditation body that validates a credential as authentic may itself be compromised or fictitious. A very common strategy employed by diploma mills is to set up fake accreditation mills to legitimize the credentials they sell.

The Federal Investigation Agency in Pakistan, has probed several instances where regulatory bodies verified fake degrees of powerful officials without due diligence \cite{hecscandal}. Recently a company investigating credentials of Chinese student applicants on behalf of prominent universities in the US had to withdraw from the project on charges that it engaged in widespread application fraud itself \cite{dipont}. A Connecticut man who sold fake degrees operated a fictitious accreditation service in parallel, the National Distance Learning Accreditation Council to validate his degrees \cite{connecticutman}. 
These practices pose highly complex challenges for employers who often have limited resources available to verify academic credentials.

\begin{figure}
\centering
\includegraphics[width=9cm]{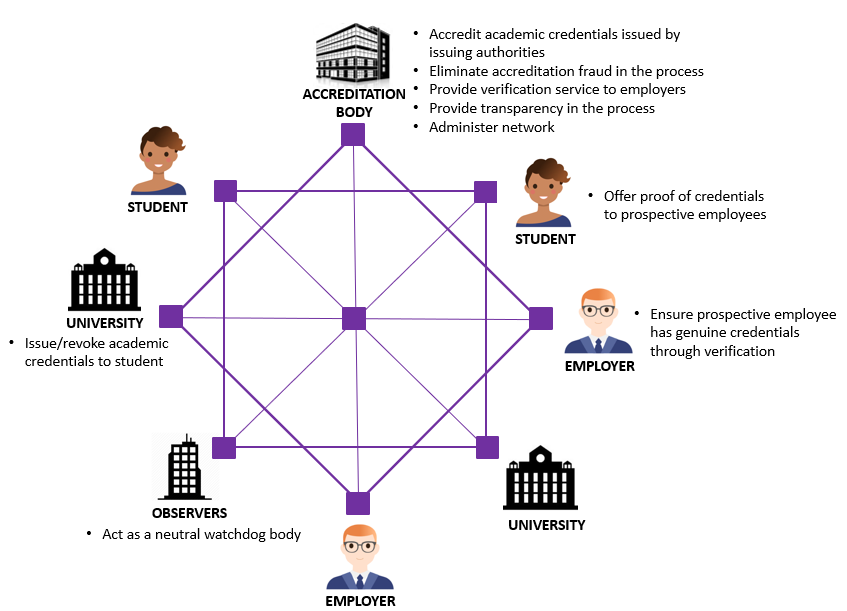}
\caption{The ecosystem for credential verification}\label{fig:ecosystem}
\end{figure}

\subsection{Blockchain and Smart Contracts}

Here we present a high-level overview of blockchain and smart contracts technology and highlight the security properties of this novel new paradigm.

A blockchain is a decentralized global ledger consisting of a continuously growing list of records, called blocks, arranged in chronological order. Users possess cryptographic credentials, namely a public/private key pair, which enables them to insert new records into this ledger. Individual blocks are coupled together using hash pointers such that data cannot be removed from the block nor added retroactively without detection by all the other parties. The synchronization and integrity of this ledger is assured by a distributed consensus protocol which periodically selects certain parties (miners) to append new blocks to the ledger.

The notion of smart contracts builds on this paradigm by envisioning the shared ledger as a public memory space, thereby extending the notion of verifiability from data to computations. The earliest cryptocurrencies, including Bitcoin, introduced limited support for scripts to govern the handling of transactions. Later platforms, notably Ethereum, provide powerful support to users to embed complex application logic into the ledger.

Various developers are using smart contracts to build sophisticated solutions which address real-world solutions in a secure and decentralized manner. These include applications such as trading securities and derivatives \cite{secder}, optimize supply chains \cite{supplychain}, remote healthcare \cite{rh}, and prediction markets \cite{predmark}, etc.

Our credential verification solution uses a permissioned blockchain, i.e. one where peers in the ecosystem are assigned different roles and privileges. Smart contracts define the rules according to which credentials may be revoked.

\subsection{Ecosystem and Threat Model}


The ecosystem for credential verification typically consist of various parties as depicted in Fig.~\ref{fig:ecosystem}.

\textbf{Universities} award academic credentials in a field of study. A \textbf{credential} in this context refers to a diploma, degree, or certificate issued by a university to a \textbf{student} in lieu of successful completion of the requirements of a certain educational program. These credentials serve as proofs of qualification, considered essential for most jobs and various other life opportunities. Therefore, there is a considerable need among \textbf{employers} and others to ensure that degrees belonging to their workers are genuine.

\textbf{Accreditation bodies} are national-level or private entities which undertake the task of verifying and validating academic credentials issued by educational institutions. Certain accreditation authorities also issue certified translations of credentials or prepare equivalence certificates. In our setup, accreditation bodies are also responsible for setting up and administering the Cerberus network.

A final party to this scenario could be \textbf{observers} such as citizen groups, activists, and watchdog bodies whose key role is to audit the operations of these different parties and maintain quality checks.

In our threat model, we assume that any of these entities may be malicious. For instance, a student may try to forge a credential or purchase one from a diploma mill. A university may sell fake degrees. An accreditation body may try to accredit fake degrees. Different malicious parties may even collude. As we've noted in \S.\ref{sec:varietiesofcredentialfraud}, these threats are realistic and there are abundant examples of each.

A credential verification solution should defend against the aforementioned attacks to the greatest possible extent. Here we list certain desirable security properties of such a system:

\begin{itemize}

\item{\textbf{Authenticity} The system should serve as a secure and authentic repository which enables verification of credentials. All stakeholders should be able to view and audit the inner workings of this system and maintain checks on the behavior of other parties.  Moreover, the system should integrate with existing credential management infrastructure.}

\item{\textbf{Resilience:} In the ideal case, we should be able to detect credential fraud if there is at least one honest party participating on the blockchain network.}

\item{\textbf{Privacy Preservation:} The system should not leak any data regarding students' credentials or personal information to third parties (such as employers) beyond any information the students may choose to reveal themselves. This includes student identities, grade transcripts, degree status, etc.}

\end{itemize}

We also list here certain other features for a credential verification system that are desirable from a usability and efficiency perspective:

\begin{itemize}
\item{\textbf{Real-time Online Verification:} A prominent property of blockchain solutions is disintermediation, i.e. decoupling the need for trusted centralized parties, and thereby immensely speeding up operations. The credential verification process is typically cumbersome and involves paperwork and communication with the awarding university. However, an online verification solution operating on top of a permissioned blockchain should enable real-time degree verification for users.}

\item{\textbf{Third-party Verification:} The system should enable third parties to directly and independently verify credentials of a user without relying on intermediaries.}

\item{\textbf{Selective Disclosure:} Users should be able to verify individual credential details in a piecemeal manner. This would enable students to share selective details with different parties. For instance, Alice could print her primary credential details on her curriculum vitae and on job applications, without revealing any other potentially sensitive data, such as national identification number or grades transcript. These could be disclosed in other situations that necessitate it.}

\item {\textbf{Usability:} The system should be easy to use and not require significant technical sophistication on the part of users. Multiple studies also note that users face difficulty storing and handling cryptographic credentials \cite{usability}\cite{usability1}.}

\item{\textbf{Revocation:} If a credential is revoked, the revocation information should be efficiently and quickly disseminated to all stakeholders, without any room for ambiguity.}
\end{itemize}

Credential verification solutions in certain countries are already adopting many of these features. Centralized online verification systems have recently been rolled out to combat widespread credential fraud in India \cite{digiloc} and Malaysia \cite{malaysia}. In later sections, we discuss how popular technologies such as smart phones and QR codes can be leveraged to further facilitate user experience.


\section{The Proposed System: Cerberus}

\begin{figure}
\centering
\includegraphics[width=9cm]{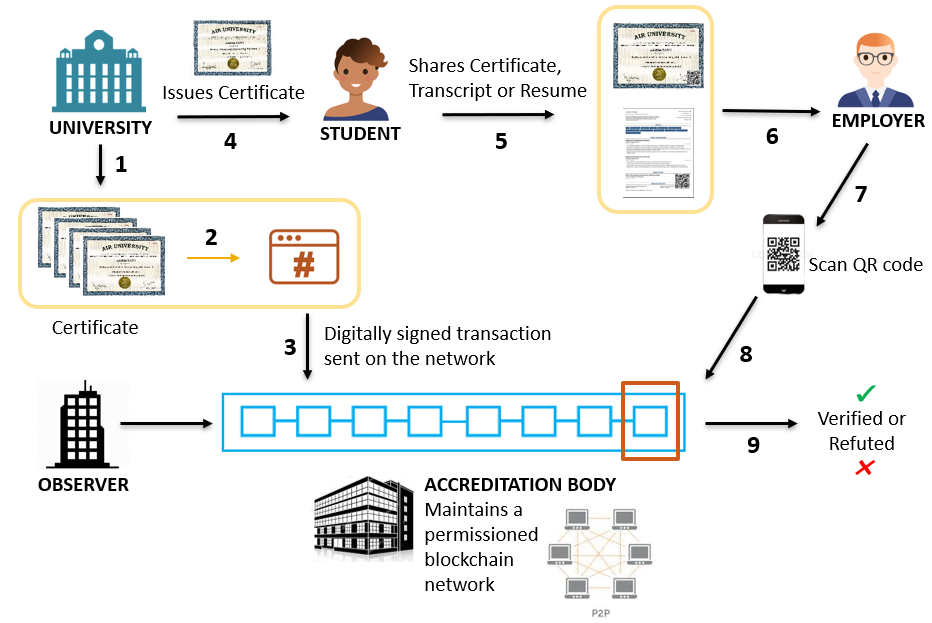}
\caption{System Overview}\label{fig:credentiallifecycle}
\end{figure}

In this section we describe the inner workings of our proposed solution. We start with a high-level overview of the key steps in the life cycle of a \textbf{credential}. The process is depicted in Fig.~\ref{fig:credentiallifecycle}.

An \textbf{accreditation body} operates and maintains a permissioned blockchain network in partnership with multiple parties including \textbf{universities} and \textbf{observer} entities, such as activist organizations and citizen groups. The \textbf{universities} circulate transactions containing validation information for credentials they issue to \textbf{students}. The \textbf{accreditation body} periodically collects these transactions and aggregates them into blocks which are then added to the blockchain. \textbf{Observer} parties audit every process in the system and maintain checks on integrity.

When a student, Alice, graduates, the \textbf{university} issues her a physical certificate of her academic degree and her transcript. The university administration also creates a transaction containing a digital fingerprint of Alice's \textbf{credential} details (alongside that of other graduating students). This transaction is digitally signed by the \textbf{university} \textbf{registrar} and propagated on the Cerberus network, where it is verified by nodes belonging to the \textbf{accreditation body}, mined into a block and added to the blockchain. This step corresponds to the issuance and accreditation of a \textbf{credential}.

The physical degree certificate issued to Alice by the \textbf{university} incorporates a QR code, for \textbf{credential} verification, printed on the front which facilitates verification of the information on her certificate.  If an \textbf{employer}, Bob, wants to check if Alice's degree is genuine, he can scan the code with a smartphone app or use a web portal. These services retrieve validating information from the Cerberus network in real-time, verify the code, and inform Bob of the degree's authenticity. Alice can also print the credential verification code on her resume alongside her educational details so that any third party can verify her degree independently of the physical certificate.

If the \textbf{university} were to revoke Alice's degree, it would circulate a revocation transaction which the \textbf{accreditation body} would verify and add to the blockchain. Any party that checks Alice's credential verification code subsequently would be informed that the degree has been revoked (as explained in detail below).

In the detailed solution specification which follows we also describe additional features whereby additional data may selectively disclosed and authenticated using the blockchain (such as Alice's identity, the contents of her educational transcript, etc.). Solution developers can easily adapt these general techniques to different types of data as per their requirements.

\subsection{Administering the Network}

The \textbf{accreditation body} is responsible for initial configuration of the network and to maintain the blockchain and update the network as participants change over time. The \textbf{accreditation body} has to deploy multiple nodes itself to distribute and secure the network. These nodes should ideally be geographically spaced and secured as per industry standards. 

The \textbf{accreditation body} also manages how other parties access and use the network. When a new \textbf{university} is listed, the \textbf{accreditation body} adds it to the network and certifies it's cryptographic keys. A \textbf{university} may also be removed from the network for various reasons (e.g. in case of fraud or if quality standards are not met) in which case it's address and keys are added to a blacklist. Keys that have been lost or compromised are also added to the blacklist and new keys are generated to replace them. The \textbf{accreditation body} also provides access to third party \textbf{observers} to audit the network.

The \textbf{accreditation body} may also update the network as per users' requirements by introducing new contracts, roles, and privileges to cater to evolving policies or changes in the ecosystem. 


\subsection{Issuance of the Credential}

When an academic session concludes, the \textbf{university} prepares degree certificates for students in the graduating batch. Issuing the \textbf{credential} comprises three key steps as follows:

\paragraph{\textbf{Preparing the Credential}}
Our solution enables verification of two sets of data: the first is data pertaining to the academic degree (denoted as \emph{degree\_info}), and second is more detailed data relating to the identity of the \textbf{student} and contents of her transcript (denoted as \emph{id/transcript\_info}). The \textbf{student} can choose to disclose these data items selectively for verification. For instance, she can publicly circulate details of her degree on her resume or her social media profile, whereas transcript and detailed identity details may only be required by some \textbf{employers}.

The first data set, \emph{degree\_info}, typically contains the following information:
\begin{itemize}
 \item   name of the Student
 \item   serial number of the degree
 \item   title of the degree/program
 \item   year the degree is awarded
 \item   name of the University
\end{itemize}

The second data set, \emph{id/transcript\_info}, consists of the following data items:
\begin{itemize}
 \item   details of student's identity document (such as drivers license, citizen card, etc.). This can even be just a personal identification number.
 \item   course codes, Titles, and credit hours for the study program
 \item   grades earned by the student
 \item   Grade Point Average and Cumulative Grade Point Average earned by the student
\end{itemize}

The items in both data sets are individually concatenated and uniquely fingerprinted using a hash function, \emph{H()}, forming the \emph{student-info} that constitutes a leaf. Popular hash functions, such as SHA2 or MD5 may be used for this purpose. The fingerprints are then input into a Merkle tree. Merkle trees, first proposed by Ralph Merkle, are hash-based data structures that allow authentication of data sets by computing a message digest (or \emph{root}) over the data items using hash-and-concatenate operations to build a tree structure encompassing the entire set \cite{merkle}. This arrangement significantly reduces the amount of verification data that needs to be put on the blockchain to verify large data sets.

\begin{figure}
\centering
\includegraphics[width=9cm]{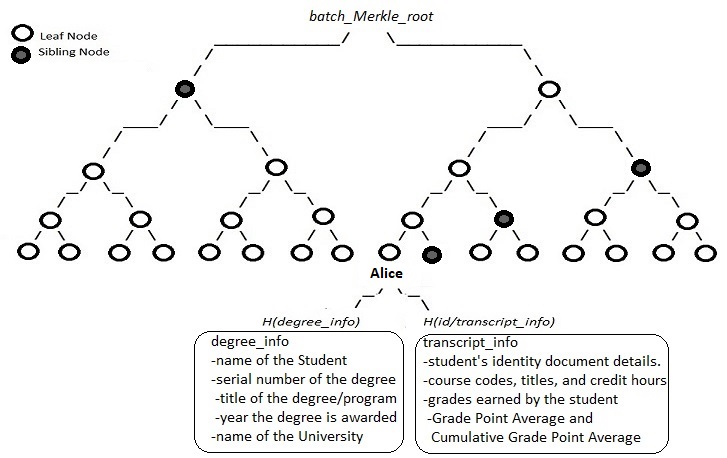}
\caption{Preparing the Credential}\label{fig:merkletree}
\end{figure}

In our solution, the tree is composed as depicted in Fig.~\ref{fig:merkletree}. Each leaf node represents a \textbf{student's} data, and the root of the tree is computed over the entire graduating batch of \textbf{students}\footnote{Merkle trees are binary, and therefore, if there are an odd number of students, the last hash will be duplicated once to create an even number of leaf nodes.}. This Merkle root, denoted the \emph{batch\_Merkle\_root}, can now be used to authenticate all data items in the original set, i.e. the degree, identity, and transcript information for all \textbf{students} in the batch. 

\paragraph{\textbf{Registering the Credential}}

The next step is to record the \emph{batch\_Merkle\_root} computed by the \textbf{university}, on to the blockchain.

The \textbf{registrar} at the \textbf{university} creates a Cerberus transaction, addressed to the \textbf{accreditation body}, embeds the \emph{batch\_Merkle\_root} in the data field of the transaction, signs the transaction with its private key, and circulates it on the Cerberus network where it is received by every party. As per the consensus rules, nodes belonging to the \textbf{accreditation body} check incoming transactions for correctness, verify the signatures, collect the transactions into a block and append it to the blockchain.

As we noted earlier in \S.\ref{sec:varietiesofcredentialfraud}, \textbf{university staff} themselves have been known to insert fake information into the student record. In our scheme, \textbf{universities} can institute further checks against insider malfeasance by deploying schemes like multi-signature wallets. This strategy requires the cooperation of additional personnel in different departments within the \textbf{university} to successfully create a valid transaction, for instance, the examination departments for the students' schools as well as the \textbf{university registrar}.

The insertion of the transaction in the blockchain serves as to accredit the \textbf{student credentials} with the \textbf{accreditation body}.

\paragraph{\textbf{Issuance of the Physical Certificates}}

\begin{figure}
\centering
\includegraphics[width=9cm]{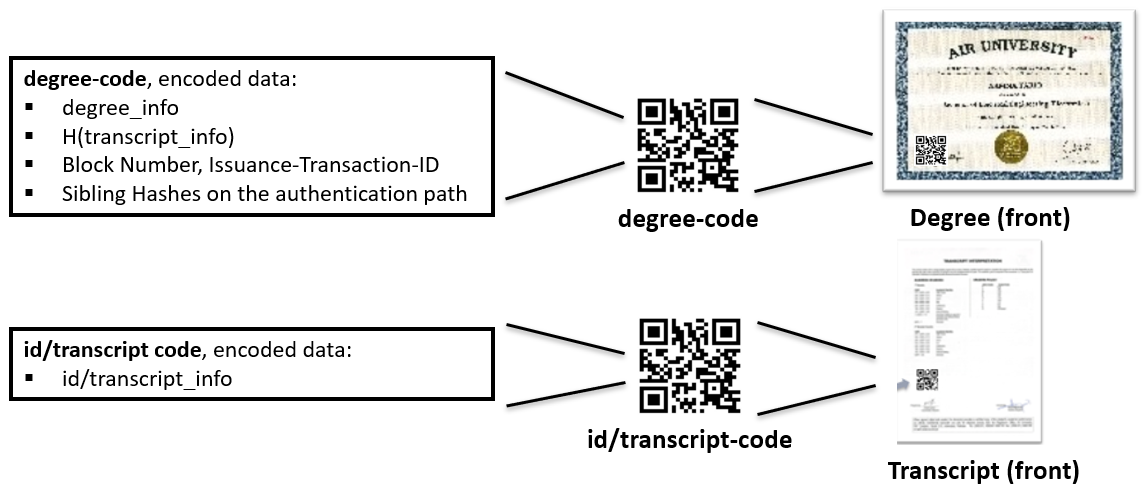}
\caption{Issuance of Physical Certificates: degree-code and id/transcript-code}\label{fig:issuephysical}
\end{figure}

The physical certificates of the \textbf{credentials} and transcripts contain printed QR codes, enabling users to validate the information using the \emph{batch\_Merkle\_root} on the blockchain. 

Merkle trees have a useful property in that individual data items in the original set over which the tree is computed may be verified independently of the other data items if the `authentication path' for the individual data items are available, i.e. those sibling nodes in the tree that share the same parent, on the path from the data item to the root. As an example, in Fig.~\ref{fig:merkletree}, Alice's degree can be authenticated by hashing the contents of the degree, concatenating a hash of her transcript and identity documents, and then reconstructing the path up to the root, using the sibling nodes \footnote{For illustration purposes, the figure shows, a  Merkle tree of height 4 and the sibling hashes on the authentication path are also 4. The number of sibling hashes on the authentication path, are always equal to the height of the Merkle tree} (shaded in Fig.~\ref{fig:merkletree}), and then verifying that the signature on the transaction belongs to the \textbf{university}. All the information that a user requires to undertake this process is embedded in the QR codes imprinted on the degree certificate. We describe these next.

The first QR code, denoted \emph{degree\_code}, is used to verify the contents of the degree certificate, and encodes the following data:
 \begin{itemize}
\item{\emph{degree\_info}, i.e. text of student's name, serial number of the degree, title of degree/program, year of the award, and name of the university}
\item{\emph{H(id/transcript\_info)}, i.e. a hash fingerprint of the data pertaining to the student's identity and transcript details, which is also necessary to reconstruct the authentication path for the \emph{degree\_info}}
\item{the \emph{block-number} and \emph{issuance-transaction-ID} for the transaction that validates the specific degree}
\item{the complete authentication path for the particular degree in the Merkle tree, i.e. all the \emph{sibling-hashes} on the path leading up to the tree root}
\end{itemize}

The second QR code, denoted \emph{id/transcript\_code}, can be used to verify the student's identity and the contents of her transcript, and encodes \emph{id/transcript\_info}, i.e. contents of student's identity document or identification number, and complete contents of her transcript. This code can be printed on the student's degree or transcript as per requirements.

We consider QR code specifications in detail in \S.~\ref{sec:prototype}. Next we describe the verification process for students and \textbf{employers}.

\begin{figure}
\centering
\includegraphics[width=9cm]{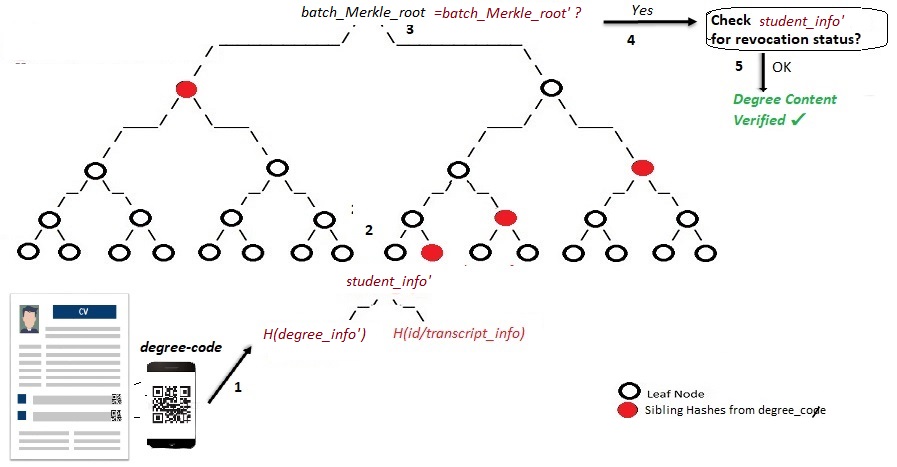}
\caption{Degree content Verification}\label{fig:Degree and Transcript Content Verification}
\end{figure}

\subsection{Verification of the Credential}

The verification process for contents of the degree and of the the identity and transcript data is very similar.

Contents of the degree are verified  by scanning the \emph{degree\_code} using a verification app, imprinted on the physical copy of the degree or if embedded into a resume. The step-wise verification process is depicted in Fig.~\ref{fig:Degree and Transcript Content Verification}.

First, the user's app computes \emph{degree\_info'} using the plaintext degree contents information in the \emph{degree\_code}. \emph{degree\_info'} and \emph{id/transcript\_info} are then concatenated and hashed. The result is then repeatedly concatenated with the appropriate sibling nodes and hashed to reconstruct Merkle tree, i.e. \emph{batch\_Merkle\_root'}. The app then queries the Cerberus blockchain (using the block number and transaction ID in \emph{degree\_code}) and checks if \emph{batch\_Merkle\_root} matches \emph{batch\_Merkle\_root'} which it has computed. If the match is positive, the contents of the degree are successfully verified.


Transcript contents are verified in a similar manner: 

The \textbf{id/transcript\_code} is scanned, and a \textbf{id/transcript\_info'} generated by the app by hashing the plaintext transcript data in the code.Then the verifier manually enters the student's identity document number. These two items are concatenated and hashed to obtain a \emph{H(id/transcript\_info'}). If this regenerated \emph{H(id/transcript\_info'} matches the \emph{H(id/transcript\_info)} (used earlier when verifying the contents of the degree) then the transcript content is also verified.

\subsection{Revoking a Credential}
Now that we've discussed degree issuance and verification, we are in a position to describe the revocation process. In certain cases, universities or accreditation bodies may choose to revoke degrees. Cerberus implements efficient on-chain multiparty revocation using smart contracts. The revocation process involves two smart contracts, the \emph{Rules Engine} which defines the precise rules according to which credentials may be revoked, Figure ~\ref{fig:revrules} and the \emph{Implementation Engine}, which enforces these rules. These contracts are summarized in Fig.~\ref{fig:revcredential}.
The addresses of these contracts are explicitly embedded in the original transaction made by the university to register the credential as described earlier. 

\begin{figure}
\begin{mdframed}  
\setlength{\parindent}{5ex}

\noindent {\fontfamily{qhv}\selectfont Contract Rules Engine}

\noindent \newline
\noindent  \textbf{Inputs:} {\fontfamily{qhv}\selectfont A${}_{n}$$\mathrm{\{}$$\mathrm{\}}$}

\noindent \textbf{Revoking Authority-List (A${}_{n}$):}

\noindent On receiving (A)

\noindent \textbf{if} 
{\fontfamily{qhv}\selectfont( E.address \textbf{$\boldsymbol{\mathrm{\in }}$ }A${}_{n}$$\mathrm{\{}$$\mathrm{\}}$):

\emph{A${}_{n}$$\{$$\}$ := A U A${}_{n}$$\{$$\}$}}

\noindent \textbf{else}
\newline
\indent {\fontfamily{qhv}\selectfont A $\in $ An$\{$$\}$ , has x 2 public keys}

\noindent \textbf{Rules (A.address):}

\noindent \textbf{if} 
{\fontfamily{qhv}\selectfont( A \textbf{$\boldsymbol{\mathrm{\in }}$} A${}_{n}$$\mathrm{\{}$$\mathrm{\}}$ \& A has not previously revoked the same document):

\indent Revoke count := Revoke count ++

\emph{return (true)}}

\noindent \textbf{else} 
\newline \indent {\fontfamily{qhv}\selectfont {Revoke count := Revoke count}}


\noindent \newline A${}_{n}$$\mathrm{\{}$$\mathrm{\}}$ = Revoking Authority-List

\noindent A = Revoking Authority

\noindent E = any Entity in the Ecosystem

\end{mdframed}
\caption{Rules Engine}
\label{fig:revrules}
\end{figure}

\paragraph{\emph{\textbf{Rules Engine}}}
\begin{itemize}
\item{Only predefined nodes (belonging to the accreditation body and/or university) can execute this contract and participate in the revocation procedure. These nodes are listed in an \emph{Authority-List}.}
\item{Revocation must be approved by at least 2 nodes, in the Authority-List.}
\item{The university and the accreditation body each have two pairs of keys. Revocation may be undertaken individually by the university or the accreditation body, or by both.}
\item{A node in the Authority-List can only sign the revocation of a certificate once.}
\item{Batch revocation is also possible in case the degree status of an entire batch of students needs to be revoked (may happened if the \textbf{university} fails to meet quality assurance criteria or if any irregularities are discovered by the \textbf{university} or \textbf{accreditation body}).}
\end{itemize}

To revoke a credential, a node in the Authority-List initiates the revocation procedure, by invoking the Implementation Engine and providing the \emph{student\_info'}. In case of batch revocation, \emph{(batch\_Merkle\_root)} is provided.

\paragraph{\emph{\textbf{Implementation Engine}}}
\begin{itemize}
\item{This contract verifies that \emph{student\_info'} matches one of the existing certificates on the blockchain.}
\item{The contract then verifies that the node is listed in the Authority-List and issues a \emph{process-hash} for the certificate, and increments a revocation counter.}
\item{A second node in the Authority-List, using the \emph{process-hash} generated in the previous step, also invokes the \emph{Implementation Engine}.}
\item{The \emph{Implementation Engine} after verifying that the calling node is listed in the Authority-List, increments the revoke count again. It now stores \emph{student\_info} in a Revoke-List.}
\end{itemize}

The verification process for the degree needs to be slightly modified to check for revocation status. In this case, the user's smartphone app invokes the \emph{Implementation Engine} present in the transaction. This contract checks if \textbf{student\_info'} is listed in the Revocation-list. If so, then it declares the \textbf{credential} as revoked.

\begin{figure}
\begin{mdframed}
\setlength{\parindent}{5ex}

\noindent {\fontfamily{qhv}\selectfont
\noindent Contract Implementation Engine\textbf{ }}
\noindent \newline

\noindent \textbf{Inputs:} {\fontfamily{qhv}\selectfont D${}_{H}$, P${}_{H}$}

\noindent {\fontfamily{qhv}\selectfont \noindent r = Rules Engine (adressRulesEngine)}

\noindent \textbf{Revoke Document (D${}_{H}$):}

\noindent {\fontfamily{qhv}\selectfont \emph{call r.Rules (A.address)}}

\noindent \textbf{if} {\fontfamily{qhv}\selectfont (r.Rules (A.address)=true): 
\newline \indent \emph{return (PH)}}

\noindent \textbf{else}

\indent {\fontfamily{qhv}\selectfont \emph{terminate process}}

\noindent \textbf{Confirm Revocation (P${}_{H}$):}

\noindent {\fontfamily{qhv}\selectfont \emph{call r.Rules (A.address)}}

\noindent \textbf{if}
{\fontfamily{qhv}\selectfont (r.Rules (A.address)=true \& Revoke count = requiredCount):}
\newline \indent {\fontfamily{qhv}\selectfont \emph{call Revoke List (DH)}}

\noindent \textbf{else} 
\newline \indent {\fontfamily{qhv}\selectfont \emph{terminate process}}

\noindent \textbf{Revoke-List (D${}_{H}$):}

\noindent {\fontfamily{qhv}\selectfont Revoke-List$\mathrm{\{}$$\mathrm{\}}$ := D${}_{H\ }$U Revoke List$\mathrm{\{}$$\mathrm{\}}$}

\noindent \textbf{Output:} {\fontfamily{qhv}\selectfont {''PH'', Revoke-List}}

\noindent \newline
\noindent D${}_{H}$ = Hash of Document
\noindent P${}_{H}$ = Process Hash
\noindent requiredCount = revocation count required for successful revocation
\end{mdframed}
\caption{Implementation Engine}
    \label{fig:revcredential}
\end{figure}

\section{Prototype}
\label{sec:prototype}
The Accreditation Body and University nodes were run on desktop computers with the following specifications:  CPU: Intel Core i7-3517U @ 1.90 GHz, Physical Memory: 4 GB DDR3 1600 MHz,  OS: Ubuntu 13.04.e 

The Students, Employers and Observers, can either use, a desktop computers, a laptop or even a smart phone. For our prototype we used one of each with the following specifications 

\begin{itemize}
    \item \textbf{Desktop} CPU: Intel Core i7-3517U @ 1.90 GHz, Physical Memory: 4 GB DDR3,  OS: Ubuntu 13.04.e
    \item \textbf{Laptop} CPU: Intel Core i3-7100U @ 2.40 GHz, Physical Memory: 8 GB DDR3,  OS: Microsoft Windows 10 Pro 10.0.17134 
    \item \textbf{Smartphone} CPU: 1.2GHz dual-core Qualcomm Snapdragon 410, Physical Memory: 1 GB , OS: Android 4.4.4 (KitKat); Sense UI, Main Camera: 8 MP
\end{itemize} 

We implemented our protoype on Parity, an Ethereum Client, Version 1.10.4-stable. Parity claims to be the fastest and most advanced Ethereum client \cite{parity}. Parity supports a private blockchain network configuration, through a Proof-of-Authority consensus engine. Proof-of-Authority, a replacement for Proof-of-Work, uses a set authorities - nodes that are explicitly allowed to create new blocks and secure the blockchain. The transactions to become permanent record and included in the blockhain have to be signed off by the predefined Authority List. This makes it easier to maintain a private chain and keep the block issuers accountable. Ethereum Virtual Machine (EVM) is the runtime environment provided by Ethereum for the efficient execution of smart-contracts. Solidity Version 4.24, was used to code the smart contracts.

The data for all the students in a batch is input and converted to a JSON format file. These JSON files, are input to the Python program for data preparation. This program deploys SHA256 where ever a hash function is required, such as for fingerprinting the data, sibling hashes and merkle roots computation etc. 

The required data for each QR code is input into an open-source QR code generator to create the QR codes to be printed on the credentials. Also an open source QR code reader was embedded into our Android app, and website interface. 

A QR (Quick Response) Code consists of black modules arranged in a square grid on a white background, that offer a unique representation of data and can be read by an imaging device. There are various types of QR codes available, distinguished by the total number of black modules and the number of modules per unit area referred to as module density. The earliest and smallest QR code called version 1 was a 21X21 grid, while each subsequent version increases by 4x4. Other distinguishing features include the level of error correction level achieved (through redundancy of data), the size of the QR Code, scanning distance, light and angle, and camera quality. It can encode numeric, alphanumeric and byte data. \cite{qr}. Each hash is 32 bytes long.

The amount of data in \emph{degree\_code} and \emph{id/transcript\_code} (for a batch of 100 students) is estimated to be around 800 alphanumeric characters, for which Table ~\ref{tab:QRtable} shows the QR code specification and size. The equation for calculating the minimum size of the QR code is as follows: Minimum Size=(Scanning Distance/Distance Factor)x(Data Density/25) \cite{qr}. For our calculation the Distance Factor =10 was assigned. 

\begin{table}[htbp]
  \centering

    \begin{tabular}{cccccc}

    \multirow{2}[2]{*}{Students} & \multirow{2}[2]{*}{Merkle Tree} & \multirow{2}[2]{*}{Sibling} & \multirow{2}[2]{*}{degree\_code} & \multicolumn{1}{c}{\multirow{2}[2]{*}{ Scanning}} & \multirow{2}[2]{*}{QR Code } \\
    
    \multirow{2}[2]{*}{in batch} & \multirow{2}[2]{*}{Height} & \multirow{2}[2]{*}{Hashes} & \multirow{2}[2]{*}{Size} & \multicolumn{1}{c}{\multirow{2}[2]{*}{ Distance}} & \multirow{2}[2]{*}{Size} \\
          &       &       &       & \multicolumn{1}{c}{} &  \\
          
     \multirow{2}[2]{*}{} & \multirow{2}[2]{*}{} & \multirow{2}[2]{*}{(bytes)} & \multirow{2}[2]{*}{(bytes)} & \multicolumn{1}{c}{\multirow{2}[2]{*}{ }} & \multirow{2}[2]{*}{(incheS)} \\
          &       &       &       & \multicolumn{1}{c}{} &  \\
          
    \midrule
    \multirow{2}[4]{*}{50} & \multirow{2}[4]{*}{6} & \multirow{2}[4]{*}{192 bytes} & \multirow{2}[4]{*}{406 bytes} & 2’’   & 0.808'' \\
         &       &       &       & 4’’   & 1.616'' \\
    \midrule
    \multirow{2}[4]{*}{100} & \multirow{2}[4]{*}{7} & \multirow{2}[4]{*}{224 bytes} & \multirow{2}[4]{*}{438 bytes} & 2’’   & 0.832'' \\
         &       &       &       & 4’’   & 1.664'' \\
    \midrule
    \multirow{2}[4]{*}{200} & \multirow{2}[4]{*}{8} & \multirow{2}[4]{*}{256 bytes} & \multirow{2}[4]{*}{470 bytes} & 2’’   & 0.904'' \\
          &       &       &       & 4’’   & 1.808'' \\
    \midrule
    \multirow{2}[4]{*}{500} & \multirow{2}[4]{*}{9} & \multirow{2}[4]{*}{288 bytes} & \multirow{2}[4]{*}{502 bytes } & 2’’   & 0.904'' \\
         &       &       &       & 4’’   & 1.808'' \\
    \midrule
    \multirow{2}[4]{*}{1000} & \multirow{2}[4]{*}{10} & \multirow{2}[4]{*}{320 bytes} & \multirow{2}[4]{*}{534 bytes} & 2’’   & 0.936'' \\
         &       &       &       & 4’’   & 1.872'' \\
    \midrule
    \multirow{2}[4]{*}{2000} & \multirow{2}[4]{*}{11} & \multirow{2}[4]{*}{352 bytes} & \multirow{2}[4]{*}{566 bytes} & 2’’   & 0.968'' \\
        &       &       &       & 4’’   & 1.965'' \\
    \midrule
    \multirow{2}[4]{*}{4000} & \multirow{2}[4]{*}{12} & \multirow{2}[4]{*}{384 bytes} & \multirow{2}[4]{*}{596 bytes} & 2’’   & 0.968'' \\
         &       &       &       & 4’’   & 1.965'' \\
    
    \end{tabular}%
    \caption{Batch Size \& QR Code Specification}
  \label{tab:QRtable}%
\end{table}%

\section{Discussion} \label{Discussion}

In this section we discuss how our scheme counteracts common types of credential fraud described earlier in \S.\ref{sec:varietiesofcredentialfraud} and highlight its novel properties.

Cerberus safeguards Academic Credentials against modification and manipulation, undue access, while making them available to legitimate entities seeking to verify the existence of a record, and surety that it has not been altered or been revoked since. We separately reason how Cerberus prevents the occurrence of each type of Credential Fraud as defined previously in the \ref{sec:varietiesofcredentialfraud}.

The permissioned blockchain architecture ensures strict separation of roles and privileges among the participating entities, preserves the integrity and chronological ordering of the credential record, and enables stakeholders to efficiently audit the operations of the system. If even one party on the network is honest, it can detect suspicious behavior by other entities and raise the alarm.

Counterfeit credentials and altered documents can easily be detected using this system. A fake credential will not have a corresponding record in the blockchain, whereas alterations of an original document will result in a different credential fingerprint which will not match the one on the blockchain.

However, our student, Alice, may bribe administrative staff in the university to falsify a credential on her behalf. Cerberus addresses this issue in two ways: first, due to the append-only property of the blockchain, even universities and accreditation bodies cannot insert backdated records into the ledger. Rewriting entire blocks to retroactively add data would require the university to actively collude with the accreditation authority, and this activity would be visible to all other parties on the blockchain.

At best, Alice, can influence staff members to insert a falsified record for her in the next upcoming batch on the blockchain. This concern may also be mitigated if the university were to define multi-signature policies for registering credentials on the blockchain. In this case, for example, creating a transaction would require the university registrar as well as administrative staff in the relevant department within the university to independently vet and sign the transaction. Alice would therefore have to bribe multiple disassociated parties to procure a fake credential, which will hopefully require considerably more effort.

Likewise, this solution would address the problem of diploma mills. Any new university wishing to join the network would first have to be thoroughly vetted by the accreditation body which would certify the university's public key, thereby allowing the university to make transactions. Non-approved entities such as fake universities, by default, would not be able to publish any data on the blockchain.

Combating fake accreditation agencies is more complex, as a resourceful attacker might set up an entire blockchain network himself to validate his fake credentials. In the context of a single nationwide or regional accreditation service, there should ideally be no confusion regarding which accreditation body is legitimate. In areas where accreditation services are privatized, the situation can perhaps be mitigated by requiring accreditation bodies to include legitimating information on the blockchain itself, such as a license to operate or a statement of authorization from relevant government ministries or reputed education watchdog bodies. The presence of reputed activist and watchdog bodies which have joined the network as observers will add to the legitimacy of the accreditation body. 

Cerberus has other fundamental advantages over legacy credential verification solutions. First and foremost, the blockchain is integrated into the credential issuance process itself, thereby enabling \emph{accreditation-by-default} in a sense. The verification process consists of only a simple lookup on the blockchain. This process dispenses with time-consuming and cumbersome paperwork that is typical in existing systems and can be undertaken in real-time using computers and smartphones. Students and employers can verify the provenance and authenticity of credentials themselves and at their convenience without relying on third parties or requiring specialized technical skills or maintaining cryptographic keys.

Privacy of student data is maintained since no actual data is put on the blockchain but only data fingerprints. These are computed using one-way hash functions thereby disassociating the process of validation of the data from the data itself. No third party is therefore able to deduce students' degree details or information from the blockchain alone without also accessing the authentication path data that is printed on the physical certificates. Cerberus enables the student to exercise personal control over this data and share it with various parties to undertake verification.

Selective disclosure is also facilitated in that the student can share an individual data item (degree data or identification data and transcript contents) and it's authentication path with a third party without having to reveal to them all the original data items.

Cerberus also significantly improves on state-of-the-art blockchain-based credential verification solutions proposed in the literature. We discuss these next.

\section{Related Work}

 \begin{table*}
  \centering
\begin{tabular}{ccccccccccc}
\textbf{Scheme} & \multicolumn{3}{c}{\textbf{System Features}} & \multicolumn{4}{c}{\textbf{Security Features}} & \multicolumn{3}{c}{\textbf{Usability}} \\
\midrule
      & \begin{turn}{45}\textit{Accreditation}\end{turn} & \begin{turn}{45}\textit{Verification}\end{turn} & \begin{turn}{45}\textit{Revocation}\end{turn} & \multicolumn{1}{p{2.5em}}{\begin{turn}{45}\textit{Counterfeit\newline{}Protection}\end{turn}} & \begin{turn}{45}\textit{Privacy}\end{turn} & \multicolumn{1}{p{2.5em}}{\begin{turn}{45}\textit{Selective\newline{} Disclosure }\end{turn}} & \begin{turn}{45}\textit{Transparency}\end{turn} & \multicolumn{1}{p{2.5em}}{\begin{turn}{45}\textit{User\newline{}Experience }\end{turn}} & \multicolumn{1}{p{2.5em}}{\begin{turn}{45}\textit{No Key\newline{}Management}\end{turn}} & \begin{turn}{45}\textit{Accessibility}\end{turn} \\
\midrule
\textbf{UNIC\cite{nicosia}} & -     & \LEFTcircle &       & \LEFTcircle & \CIRCLE & -     & \CIRCLE & \CIRCLE & \CIRCLE & \CIRCLE \\
\midrule
\textbf{Blockcerts*\cite{blockcerts}} & -     & \CIRCLE & \LEFTcircle & \LEFTcircle & \CIRCLE & -     & \CIRCLE & \CIRCLE & - & \CIRCLE \\
\midrule
\textbf{Hypercert\cite{hypercert}} & -     & \CIRCLE & \LEFTcircle & \LEFTcircle & \CIRCLE & -     & \CIRCLE & \CIRCLE & - & \CIRCLE \\
\midrule
\textbf{Echo\textdagger\cite{echo}} & -     & \LEFTcircle &       & \LEFTcircle & -     & -     & \CIRCLE & \LEFTcircle & -     & \LEFTcircle \\
\midrule
\textbf{UZHBC\textdagger\cite{uzhbc}} & -     & \LEFTcircle &       & \LEFTcircle & \CIRCLE & -     & \CIRCLE & \CIRCLE & \CIRCLE & \CIRCLE \\
\midrule
\textbf{EduCtx\textdagger \cite{EduCTX}} & \CIRCLE & \LEFTcircle &       & \CIRCLE & \CIRCLE & -     & \LEFTcircle & \LEFTcircle & -     & \LEFTcircle \\
\midrule
\multicolumn{1}{p{7.5em}}{\textbf{Blockchain for\newline{} Education\textdagger\cite{bc4edu}}} & \CIRCLE & \CIRCLE & \LEFTcircle & \CIRCLE & \CIRCLE & -     & \CIRCLE & \CIRCLE & \CIRCLE & \CIRCLE \\
\midrule
\textbf{Cerberus} & \CIRCLE & \CIRCLE & \CIRCLE & \CIRCLE & \CIRCLE & \CIRCLE & \CIRCLE & \CIRCLE & \CIRCLE & \CIRCLE \\
\bottomrule
\end{tabular}%
    \begin{tablenotes}
    \centering
    \item[1]  \CIRCLE = provides property; \LEFTcircle = partially provides property; - = does not provide property
   \item[2] \textdagger = has academic publication; * = end-user tool available
  \end{tablenotes}
      \caption{Summary comparison of various solutions}
  \label{tab:summarycomparison}%
\end{table*}%

\textbf{Legacy credential verification procedures} typically rely on university databases and unique codes (pin numbers and hashing) and anti-counterfeiting technology to establish the credibility of a document \cite{certisafe}, \cite{uniquecodes}, \cite{parchment}. Some solutions associate user profiles and education records with a single identity \cite{badgr} \cite{mozillaopenbadges}. These systems have numerous weaknesses: there is systematic redundancy, tedious paperwork, and cumbersome processes, resulting in extensive effort, cost, and delays. There is little standardization. Moreover loopholes for fraud and crime persist, as documented in \S.~\ref{sec:varietiesofcredentialfraud}.


It is widely acknowledged that the blockchain can play a pivotal role in resolving many of these issues. Various startups and companies now offer \textbf{credential verification on the blockchain}, e.g. Appii \cite{appii}, Gradba \cite{gradba}, Aversafe \cite{aversafe}, Verify \cite{verify}, Accredible \cite{accredible}, TrueRec \cite{truerec}, and Bcdiploma \cite{bcdiploma}. Efforts are also underway to build \textbf{blockchain-backed credential verification solutions at the national level} in Malaysia \cite{malaysia2018} and India \cite{indiachain}.

Academic researchers have made significant contributions in this domain and have proposed various new features and optimizations. A pioneering effort in this regard was made by the \textbf{University of Nicosia (UNIC)} in 2015, when it became the first educational institution to issue academic certificates (for its Digital Currencies course) on the Bitcoin blockchain \cite{nicosia}. Since 2017, UNIC has issued all university diplomas on the Bitcoin blockchain. Their solution is to issue the student a digitally signed PDF file of the credential. The hash of this file is inserted in the OP\_RETURN field of a Bitcoin transaction.

Another prominent example is \textbf{Blockcerts}, an initiative of the MIT Media Lab, which similarly inserted credential verification information in the Bitcoin OP\_RETURN field and the Ethereum extraData field. Blockcerts also supported batch issuance of credentials using Merkle trees. 

Blockcerts had two key shortcomings: first, it required students to maintain cryptographic keys, and, second, it maintained credential revocation lists on a centralized website which could theoretically be compromised. To resolve the second issue, researchers proposed \textbf{Hypercerts}, a distributed and trustless credential revocation mechanism, which relies on Ethereum smart contracts and the InterPlanetary File System (IPFS), a decentralized data storage solution \cite{hypercert}.

Solutions that followed contributed further to developing scope, features, and application. For instance, \textbf{EduCTX} proposes a unified global higher education credit and grading system based on the European Credit Transfer and Accumulation System (ECTS), in which coins are transferred on the blockchain to signify academic study credits attained by students \cite{EduCTX}. This solution, built on the ARK platform, also requires students to maintain cryptographic keys.

EchoLink stores user identities and academic credentials on-chain via smart contracts for a range of blockchain platforms, including Ethereum, AntShares, Metaverse, etc. \cite{echo} It's goal is to build a professional networking and recruiting platform, providing easy access to a pool of vetted candidates. EchoLink does not have a revocation mechanism. 

\textbf{UZHBC (University of ZuricH BlockChain)} is a blockchain-based verification system, specifically for diplomas issued by the University of Zurich \cite{uzhbc}. It uses the public Ethereum blockchain and employs a smart contract for both issuance and verification functions, and accepts a PDF of the credential as input. Likewise, \textbf{Blockchain for Education} also uses the public Ethereum blockchain and smart contracts for access control and certificate management \cite{bc4edu}. The Interplanetary Filesystem supplements this system by providing users access to profile information of certification authorities.



These systems have various limitations: for instance, UNIC and UZHBC are limited in scope to their parent institution and EchoLink is only available to registered users. Most of these solutions do not incorporate accreditation bodies (with the notable exception of EduCTX and Blockchain for Education), which leaves open the issue of university staff falsifying records and the problem of diploma mills.

Most of these systems preserve privacy of student data. EchoLink is the notable exception since it is recruiting platform and offers viewing access to registered users. 

The issue of scalability is only addressed by Blockcerts, which offers a batch issuance mechanism. Other systems rely on separate transactions per student which add to transaction costs and contribute to blockchain-bloat when deployed on public blockchains.

Moreover, the significant problem of certificate revocation, is only satisfactorily addressed by two systems: Hypercerts which relies on the InterPlanetary File System network, and by Blockchain for Education which claims to use smart contracts for revocation (implementation details are not provided).

In addition, many systems, notably Blockcerts, EchoLink and EduCTX, require students and verifiers to maintain cryptographic credentials or digital identities to participate in the ecosystem and avail the verification service. This complicates the user experience as has been documented in multiple studies in the literature. 

We compare key properties of these solutions with our work in \ref{tab:summarycomparison}. Cerberus makes significant new contributions: for one, our solution adheres closely to the established ecosystem and enables an extra layer of oversight by incorporating accreditation authorities and independent observer and watchdog bodies. This is an effective check against documented real-world problems such as that of corrupt university staff and the phenomenon of diploma mills.

Cerberus relies on batch issuance and is therefore scalable and can function as a university, consortium, or national-level solution. Our revocation scheme is also unique in that it is on-chain and is a multi-party scheme, making it harder to abuse. Cerberus also maintains privacy of user data and offers students the novel facility of selective disclosure of their verification information as per requirements.



A key advantage of Cerberus over prior solutions is in terms of usability. Verification is an intuitive process which simply requires scanning QR codes. Users do not have to maintain digital identities or cryptographic credentials. QR codes are easy and intuitive to use. Students can even distribute the codes on their resumes and circulate them publicly if they choose, enabling verification without the physical credential.


Finally, Cerberus is a private and permissioned system, which, in contrast to public blockchains, enables more streamlined management and enforcement of policy and rules, and the system can be upgraded to cater to the changing requirements of the ecosystem.


\section{Future Work}

Here we describe possibilities for enhancing Cerberus and adding new features.

One straightforward addition is to link multiple credentials or qualifications undertaken by a student such that verifying the most recent one ensures that all her previous credentials are also genuine. This can be easily incorporated, by including hashes of the earlier credentials in the transaction for her recent credential. This would simplify the verification process in cases multiple qualifications in the candidate's history need to be verified. 

Second, Cerberus is a private network, and as an additional integrity check, periodic snapshots of the Cerberus blockchain can be anchored to a public blockchain, such as Bitcoin, Ethereum etc. A convenient option for the snapshot would be the hash of the latest block. This provides security against malicious forks and history revision attacks if all parties in the network collude and insert data retroactively in the blockchain \cite{bittertobetter}. Blocks may only be considered part of the canonical record if they build on top of the most recent block snapshots.

Furthermore, as blockchain-based solutions permeate management and governance structures (as has been predicted \cite{hbr}), we anticipate there will be solutions for various public records (e.g. identity-information, health, education, driving license, passport, criminal records, etc.). This may eventually give rise to overarching records management systems, which link multiple blockchains in the background as \emph{sidechains} to a main chain, enabling them to securely and efficiently exchange information and present a unified and integrated platform.

A relevant example is IndiaChain, a high-level blockchain platform with which other sidechains are expected to interface \cite{indiachain}. Various sidechain protocols have been proposed to date \cite{channels} \cite{omniledger} \cite{plasma} include prominent projects such as Rootstock \cite{rootstock} and Elements \cite{elements} that are sidechains linking to the Bitcoin network. This architecture introduces modularity in the system as well as obvious advantages of scalability and interoperability. Whereas interoperability protocols are still at a nascent stage in their development \cite{crosschain}  \cite{multiplebc}, it would be interesting to devise integration solutions for Cerberus in a unified platform and investigate new features and opportunities of such a step.
 
\section{Conclusion}

Credential fraud is a widespread and pervasive practice that undermines confidence in educational institutions, impairs social development, and involves significant economic costs. Unfortunately, legacy credential verification systems are time-consuming, costly, and cumbersome. Moreover, they are not very effective against certain widespread corrupt practices, including fraud on the part of educational institutions and accreditation bodies.

In this paper we have proposed Cerberus, a comprehensive blockchain-based solution which counters widespread instances of fraud, as well as offers dramatic improvements over legacy systems in terms of usability and efficiency. Cerberus also offers distinct benefits over existing blockchain-based solutions proposed in the literature and in industry. Our solution integrates effectively with the existing credential verification ecosystem, it includes a novel on-chain credential revocation mechanism, and does not require students or employers to maintain cryptographic credentials.

We hope this work contributes positively to ongoing and future efforts towards alleviating the phenomenon of credential fraud.


\begin{thebibliography}{}

\bibitem{socialmobility}
Haveman, Robert, and Timothy Smeeding. "The role of higher education in social mobility." The Future of children (2006): 125-150.

\bibitem{socialmobility1}
Brown, Phillip. "Education, opportunity and the prospects for social mobility." British Journal of Sociology of Education 34, no. 5-6 (2013): 678-700.

\bibitem{ecogrowth}
Hanushek, Eric A., and Ludger Woessmann. "Education and economic growth." Economics of education (2010): 60-67.

\bibitem{ecogrowth1}
Hanushek, Eric A., and Dennis D. Kimko. "Schooling, labor-force quality, and the growth of nations." American economic review 90, no. 5 (2000): 1184-1208.

\bibitem{ecogrowth2}
Coulombe, Serge, Jean-François Tremblay, and Sylvie Marchand. Literacy scores, human capital and growth across fourteen OECD countries. Ottawa: Statistics Canada, 2004.

\bibitem{schweinhart}
Schweinhart, L., and Z. Xiang. "Evidence that the High/Scope Perry Preschool program prevents adult crime." In American Society of Criminology Conference. 2003.

\bibitem{verba}
Verba, Schlozman. "Brady 1995 Verba S., Schlozman KL, Brady HE Voice and Equality." (1995).

\bibitem{reimers}
Reimers, Fernando. "Citizenship, identity and education: Examining the public purposes of schools in an age of globalization." Prospects 36, no. 3 (2006): 275-294.

\bibitem{humancapital}
Barro, Robert J. "Human capital and growth." American economic review 91, no. 2 (2001): 12-17.

\bibitem{corredu}
Cardenas-Denham, Sergio. "Corruption in Education: A Review of the Literature." PhD diss., Harvard Graduate School of Education, 2007.

\bibitem{ti}
Transparency International. Global corruption report: Education. Taylor \& Francis, 2013. Available:
https://www.transparency.org/whatwedo/publication/global\_corruption\_report\ \_education. [Accessed: 11- Sep- 2018].

\bibitem{gilles}
Grolleau, Gilles, Tarik Lakhal, and Naoufel Mzoughi. "An introduction to the economics of fake degrees." Journal of Economic Issues 42, no. 3 (2008): 673-693.

\bibitem{cats}
"Cat Gets MBA Degree - Money News Story - WCAU | Philadelphia", Web.archive.org, 2019. [Online]. Available: https://web.archive.org/web/20071026103729/http://www.nbc10.com /money/3975070/detail.html. [Accessed: 19- Feb- 2019].

\bibitem{dogs}
"Education: Sending Degrees to the Dogs", TIME.com, 2019. [Online]. Available: http://content.time.com/time/magazine/article/0,9171
,954229-1,00.html. [Accessed: 19- Feb- 2019].

\bibitem{lor}
"Lying on your resume", Monster Career Advice, 2018. [Online]. Available: https://www.monster.com/career-advice/article/lying-on-your-resume. [Accessed: 28- Feb- 2018].

\bibitem{unizulu}
D. Davis, "Unizulu professor’s murder linked to fake PhD syndicate at university", Briefly, 2019. [Online]. Available: https://briefly.co.za/20375-unizulu-professors-murder-linked-fake-phd-syndicate-university.html. [Accessed: 19- Feb- 2019].

\bibitem{ezell}
Ezell, A. (2015, November 10). The “Axact” Scam and the Big Business of Credential Fraud [webinar]. American Association of Collegiate Registrars and Admissions Officers

\bibitem{doctors}
"Doctors fake degrees score reaches 38", Thenews.com.pk, 2019. [Online]. Available: https://www.thenews.com.pk/archive/print/248189-doctors-fake-degrees-score-reaches-38. [Accessed: 19- Feb- 2019].

\bibitem{nurses}
C. Turner, "NHS consultants and nurses accused of buying fake degrees online", The Telegraph, 2019. [Online]. Available: https://www.telegraph.co.uk/education/2018/01/16/nhs-consultants-nurses-accused-buying-fake-degrees-online/. [Accessed: 19- Feb- 2019].

\bibitem{pilots}
"24 PIA pilots held fake degrees, CAA informs SC | The Express Tribune", The Express Tribune, 2019. [Online]. Available: https://tribune.com.pk/story/1706726/1-24-active-pia-pilots-fake-degrees-caa-informs-sc/. [Accessed: 19- Feb- 2019].

\bibitem{piapilots}
"Fake degree holders deserve no compassion: CJP | The Express Tribune", The Express Tribune, 2019. [Online]. Available: https://tribune.com.pk/story/1884489/1-sc-disposes-pia-fake-degrees-case/. [Accessed: 19- Feb- 2019].

\bibitem{politicians}
"Politicians, Fake Degrees and Plagiarism | Inside Higher Ed", Insidehighered.com, 2018. [Online]. Available: https://www.insidehighered.com/blogs/world-view/politicians-fake-degrees-and-plagiarism. [Accessed: 05- Dec- 2018].

\bibitem{russia}
"Breaking down Russia's culture of fake degrees", Radio National, 2016. [Online]. Available: https://www.abc.net.au/radionational/programs

/latenightlive/breaking-down-russias-culture-of-fake-degrees/7482698. [Accessed: 19- Feb- 2019].

\bibitem{jordan}
"SA MP resigns over fake degree", BBC News, 2019. [Online]. Available: https://www.bbc.com/news/world-africa-28745435. [Accessed: 19- Feb- 2019].

\bibitem{chinaqual}
"China Qualifications Verification", Chinadegrees.cn, 2019. [Online]. Available: https://www.chinadegrees.cn/en/. [Accessed: 19- Feb- 2019].

\bibitem{hecver}
"Degree Attestation System", Hec.gov.pk, 2019. [Online]. Available: http://hec.gov.pk/english/services/students/Degree\%20Attestation
\%20System/Pages/Degree-Attestation.aspx. [Accessed: 19- Feb- 2019].

\bibitem{russiaattest}
Vak.ed.gov.ru, 2019. [Online]. Available: http://vak.ed.gov.ru/87. [Accessed: 19- Feb- 2019].

\bibitem{chea}
"Home | Council for Higher Education Accreditation", Chea.org, 2019. [Online]. Available: https://www.chea.org/. [Accessed: 19- Feb- 2019].

\bibitem{teqsa}
"Tertiary Education Quality and Standards Agency", Teqsa.gov.au, 2019. [Online]. Available: https://www.teqsa.gov.au/. [Accessed: 19- Feb- 2019].

\bibitem{hague}
"Apostille Convention", En.wikipedia.org, 2018. [Online]. Available: https://en.wikipedia.org/wiki/Apostille\_Convention. [Accessed: 15- Sep- 2018].

\bibitem{whed}
(IAU), "World Higher Education Database (WHED) Portal", Whed.net, 2018. [Online]. Available: https://whed.net/home.php. [Accessed: 15- Sep- 2018].

\bibitem{independent}
R. Garner, "A third of employers never check job applicants' qualifications,", The Independent, 2014. [Online]. Available: https://www.independent.co.uk/news/education/education-news/a-third-of-employers-never-check-job-applicants-qualifications-survey-finds-9681286.html. [Accessed: 12- Sep- 2018].

\bibitem{badhire}
London, "Aversafe Meets Global Market Needs – Aversafe", Aversafe, 2018. [Online]. Available: https://blog.aversafe.com/aversafe-meets-global-market-needs-6a7489ef67fe. [Accessed: 12- Sep- 2018].

\bibitem{ripple}
"Ripple - One Frictionless Experience To Send Money Globally | Ripple", Ripple, 2019. [Online]. Available: https://ripple.com/. [Accessed: 19- Feb- 2019].

\bibitem{stellar}
"Stellar - Develop the world's new financial system", Stellar, 2019. [Online]. Available: https://www.stellar.org/. [Accessed: 19- Feb- 2019].

\bibitem{binded}
"Binded: Copyright made simple", Binded.com, 2018. [Online]. Available: https://binded.com/. [Accessed: 15- Feb- 2018]

\bibitem{ascribe}
"Artists \& Creators | ascribe", ascribe, 2018. [Online]. Available: https://www.ascribe.io/. [Accessed: 15- Feb- 2018]

\bibitem{bnotary}
"Bitcoin.com", Notary.bitcoin.com, 2018. [Online]. Available: https://notary.bitcoin.com/. [Accessed: 11- Jan- 2018]

\bibitem{proofofexistence}
"Proof of Existence", Proofofexistence.com, 2018. [Online]. Available: https://www.proofofexistence.com/. [Accessed: 05- Dec- 2018].

\bibitem{swedenland}
J. Young, "Sweden Officially Started Using Blockchain to Register Land and Properties", Cointelegraph, 2019. [Online]. Available: https://cointelegraph.com/news/sweden-officially-started-using-blockchain-to-register-land-and-properties. [Accessed: 19- Feb- 2019].

\bibitem{smartdubai}
"Blockchain | Smart Dubai", Smartdubai.ae, 2019. [Online]. Available: https://www.smartdubai.ae/initiatives/blockchain. [Accessed: 19- Feb- 2019].

\bibitem{e-Gov}
Ølnes, Svein, and Arild Jansen. "Blockchain Technology as s Support Infrastructure in e-Government." In International Conference on Electronic Government, pp. 215-227. Springer, Cham, 2017.

\bibitem{e-Chaina}
Hou, Heng. "The Application of Blockchain Technology in E-Government in China." In Computer Communication and Networks (ICCCN), 2017 26th International Conference on, pp. 1-4. IEEE, 2017.

\bibitem{medrec}
Azaria, Asaph, Ariel Ekblaw, Thiago Vieira, and Andrew Lippman. "Medrec: Using blockchain for medical data access and permission management." In Open and Big Data (OBD), International Conference on, pp. 25-30. IEEE, 2016.

\bibitem{medshare}
Xia, Qi, Emmanuel Boateng Sifah, Kwame Omono Asamoah, Jianbin Gao, Xiaojiang Du, and Mohsen Guizani. "MeDShare: Trust-Less Medical Data Sharing Among Cloud Service Providers via Blockchain." IEEE Access 5 (2017): 14757-14767.

\bibitem{e-health}
Mettler, Matthias. "Blockchain technology in healthcare: The revolution starts here." In e-Health Networking, Applications and Services (Healthcom), 2016 IEEE 18th International Conference on, pp. 1-3. IEEE, 2016.

\bibitem{clinical}
Shae, Zonyin, and Jeffrey JP Tsai. "On the Design of a Blockchain Platform for Clinical Trial and Precision Medicine." In Distributed Computing Systems (ICDCS), 2017 IEEE 37th International Conference on, pp. 1972-1980. IEEE, 2017.

\bibitem{blockfreight}
"Blockfreight, Inc. [BFT:XCPC]", Blockfreight, Inc. [BFT:XCPC], 2019. [Online]. Available: https://blockfreight.com/. [Accessed: 19- Feb- 2019].

\bibitem{everledger}
"Everledger | A Digital Global Ledger", Everledger.io, 2018. [Online]. Available: https://www.everledger.io/. [Accessed: 7- Jan- 2018]

\bibitem{agora}
"Agora", Agora, 2019. [Online]. Available: https://www.agora.vote/. [Accessed: 19- Feb- 2019].

\bibitem{followmyvote}
"The Online Voting Platform of The Future - Follow My Vote", Follow My Vote, 2019. [Online]. Available: https://followmyvote.com/. [Accessed: 19- Feb- 2019].

\bibitem{aversafe}
"Decentralized credential verification", Aversafe - Decentralized Credential Verification, 2018. [Online]. Available: https://www.aversafe.com/. [Accessed: 05- Dec- 2018].

\bibitem{bc4edu}
Kolvenbach, Sabine, Rudolf Ruland, Wolfgang Gräther, and Wolfgang Prinz. "Blockchain 4 Education." In Proceedings of 16th European Conference on Computer-Supported Cooperative Work-Panels, Posters and Demos. European Society for Socially Embedded Technologies (EUSSET), 2018.

\bibitem{EduCTX}
Turkanović, Muhamed, Marko Hölbl, Kristjan Košič, Marjan Heričko, and Aida Kamišalić. "EduCTX: A blockchain-based higher education credit platform." IEEE Access (2018).

\bibitem{blockcerts}
"Blockchain Credentials", Blockcerts, 2018. [Online]. Available: https://www.blockcerts.org/. [Accessed: 11- Jan- 2018].

\bibitem{vevo}
"Check visa details and conditions", Immi.homeaffairs.gov.au, 2019. [Online]. Available: https://immi.homeaffairs.gov.au/visas/already-have-a-visa/check-visa-details-and-conditions/check-conditions-online. [Accessed: 19- Feb- 2019].

\bibitem{ecsuk}
"Use the Employer Checking Service", GOV.UK, 2019. [Online]. Available: https://www.gov.uk/employee-immigration-employment-status. [Accessed: 19- Feb- 2019].

\bibitem{everifyus}
"Home", E-Verify, 2019. [Online]. Available: https://www.e-verify.gov/. [Accessed: 19- Feb- 2019].

\bibitem{zaretsky}
Zaretskiy, Yury. "Fake Academic Degrees in the 18th Century?." (2016).

\bibitem{credentialism}
Johnson, Creola. "Credentialism and the proliferation of fake degrees: The employer pretends to need a degree; the employee pretends to have one." Hofstra Lab. \& Emp. LJ 23 (2005): 269.

\bibitem{transnationaledu}
N. Clarke, "Understanding Transnational Education, Its Growth and Implications - WENR", WENR, 2018. [Online]. Available: https://wenr.wes.org/2012/08/wenr-august-2012-understanding-transnational-education-its-growth-and-implications. [Accessed: 11- Sep- 2018].

\bibitem{wenr}
S. Trines, "Academic Fraud, Corruption, and Implications for Credential Assessment", WENR, 2018. [Online]. Available: https://wenr.wes.org/2017/12/academic-fraud-corruption-and-implications-for-credential-assessment. [Accessed: 11- Sep- 2018].

\bibitem{scourge}
Mohamedbhai, Goolam. "The scourge of fraud and corruption in higher education." International Higher Education 84 (2016): 12-14.

\bibitem{verifilereport}
Cohen, Eyal Ben, and Rachel Winch. "Diploma and accreditation mills: New trends in credential abuse." Bedford: Verifile Accredibase (2011).

\bibitem{ezzelbook2005}
Ezell, Allen, and John Bear. Degree mills: The billion-dollar industry that has sold over a million fake diplomas. Pyr Books, 2005.

\bibitem{ezzelbook2012}
Bear, John. Degree Mills: The Billion-Dollar Industry That Has Sold Over a Million Fake Diplomas. Prometheus Books, 2012.

\bibitem{univofwales}
Henry, "University of Wales abolished after visa scandal", Telegraph.co.uk, 2018. [Online]. Available: https://www.telegraph.co.uk/education/educationnews/8843200/University-of-Wales-abolished-after-visa-scandal.html. [Accessed: 04- Dec- 2018].

\bibitem{walesstats}
G Altbach, P. (2012). Taking on corruption in international higher education. [online] University World News. Available at: https://www.universityworldnews.com/post.php?story=20120717134058780 [Accessed 28 Feb. 2019].

\bibitem{thehindu}
"Degree certificate racket thrives in Bengaluru", The Hindu, 2018. [Online]. Available: https://www.thehindu.com/news/cities/bangalore
/degree-certificate-racket-thrives-in-bengaluru/article19127959.ece?utm\_content=buffer69bc3\&utm\_medium=social\&utm\_source=twit-
ter.com\&utm\_campaign=buffer. [Accessed: 04- Dec- 2018].

\bibitem{indiamedical}
Gitanjali, B. "Academic dishonesty in Indian medical colleges." Journal of postgraduate medicine 50, no. 4 (2004): 281.

\bibitem{rediffnews}
 “Tomar’s law degree is indeed fake, Smriti’s case is different.” 9 June. Available: http://www.rediff.com/news/report/tomars-lawdegree-is-indeed-fake-smritis-case-is-different/20150609.htm
(Accessed 04- Dec- 2018).

\bibitem{cvc}
"CVC Nigeria - Committee of Vice Chancellors of Nigerian Universities.", Cvcnigeria.org, 2018. [Online]. Available: https://www.cvcnigeria.org/view.php?id=66. [Accessed: 04- Dec- 2018].

\bibitem{nytimes}
D. Schemo, "Diploma Mill Concerns Extend Beyond Fraud", Nytimes.com, 2018. [Online]. Available: https://www.nytimes.com/2008/06/29/us/29diploma.html. [Accessed: 04- Dec- 2018].

\bibitem{uganda}
Barigaba, J. (2016). Uganda: Scandal - How Ugandan Varsity Awarded 1,000 Sudan Sudanese Degrees in Months. [online] All Africa. Available at: https://allafrica.com/stories/201611290869.html [Accessed 28 Feb. 2019].

\bibitem{axact}
D. Asad, "Axact CEO, 22 others sentenced to 20 years in jail in fake degrees case", DAWN.COM, 2018. [Online]. Available: https://www.dawn.com/news/1418156. [Accessed: 04- Dec- 2018].

\bibitem{axactextort}
Cheema, U. (2018). Axact fraud case: FIA report raises questions about FIA. [online] Thenews.com.pk. Available at: https://www.thenews.com.pk/print/277439-axact-fraud-case-fia-report-raises-questions-about-fia [Accessed 28 Feb. 2019].

\bibitem{hecscandal}
"FIA probing fake degrees attestation by HEC officials", Thenews.com.pk, 2019. [Online]. Available: https://www.thenews.com.pk/print/392649-fia-probing-fake-degrees-attestation-by-hec-officials. [Accessed: 20- Feb- 2019].

\bibitem{dipont}
Stecklow, S. (2017). Exclusive: Chinese firm withdraws from U.S. effort to fight college.... [online] U.S. Available at: https://www.reuters.com/article/us-usa-college-dipont-idUSKBN1521QT [Accessed 28 Feb. 2019].

\bibitem{connecticutman}
"Connecticut Man Pleads Guilty In Multi-Million Dollar Diploma Fraud", Justice.gov, 2019. [Online]. Available: https://www.justice.gov/usao-edpa/pr/connecticut-man-pleads-guilty-multi-million-dollar-diploma-fraud. [Accessed: 20- Feb- 2019].


\bibitem{secder}
Mainelli, Michael, and Alistair Milne. "The impact and potential of blockchain on the securities transaction lifecycle." (2016).

\bibitem{supplychain}
Kim, Henry M., and Marek Laskowski. "Toward an ontology‐driven blockchain design for supply‐chain provenance." Intelligent Systems in Accounting, Finance and Management 25, no. 1 (2018): 18-27.

\bibitem{rh}
Griggs, Kristen N., Olya Ossipova, Christopher P. Kohlios, Alessandro N. Baccarini, Emily A. Howson, and Thaier Hayajneh. "Healthcare blockchain system using smart contracts for secure automated remote patient monitoring." Journal of medical systems 42, no. 7 (2018): 130.

\bibitem{predmark}
Clark, Jeremy, Joseph Bonneau, Edward W. Felten, Joshua A. Kroll, Andrew Miller, and Arvind Narayanan. "On decentralizing prediction markets and order books." In Workshop on the Economics of Information Security, State College, Pennsylvania. 2014.


\bibitem{usability} Eskandari, Shayan, Jeremy Clark, David Barrera, and Elizabeth Stobert. "A first look at the usability of bitcoin key management." arXiv preprint arXiv:1802.04351 (2018).

\bibitem{usability1}
Krombholz, Katharina, Aljosha Judmayer, Matthias Gusenbauer, and Edgar Weippl. "The other side of the coin: User experiences with bitcoin security and privacy." In International Conference on Financial Cryptography and Data Security, pp. 555-580. Springer, Berlin, Heidelberg, 2016.

\bibitem{digiloc}
"DigiLocker - Online document storage facility | National Portal of India", India.gov.in, 2018. [Online]. Available: https://www.india.gov.in/spotlight/digilocker-online-document-storage-facility. [Accessed: 04- Dec- 2018].

\bibitem{malaysia}
"Penerima Ijazah Kehormat dan Ijazah Kedoktoran di Universiti Malaysia / Honorary Degree and Doctorate (PhD) Holders at Malaysian Universities", 2018. [Online]. Available: http://waecdirect.org/. [Accessed: 04- Dec- 2018].
http://dohe.mohe.gov.my/award/

\bibitem{merkle}
Merkle, Ralph C. "A digital signature based on a conventional encryption function." In Conference on the theory and application of cryptographic techniques, pp. 369-378. Springer, Berlin, Heidelberg, 1987.

\bibitem{parity}
”Parity Documentation - Parity Technologies”, Wiki.parity.io, 2018. [Online]. Available: https://wiki.parity.io/. [Accessed: 29- Sep- 2018].

\bibitem{qr}
Qrcode.com. (2019). Information capacity and versions of QR Code | QRcode.com | DENSO WAVE. [online] Available at: https://www.qrcode.com/en/about/version.html [Accessed 25 Jun. 2019].

\bibitem{blockbench}
Dinh, Tien Tuan Anh, Ji Wang, Gang Chen, Rui Liu, Beng Chin Ooi, and Kian-Lee Tan. "Blockbench: A framework for analyzing private blockchains." In Proceedings of the 2017 ACM International Conference on Management of Data, pp. 1085-1100. ACM, 2017.

\bibitem{untangling}
Dinh, Tien Tuan Anh, Rui Liu, Meihui Zhang, Gang Chen, Beng Chin Ooi, and Ji Wang. "Untangling blockchain: A data processing view of blockchain systems." IEEE Transactions on Knowledge and Data Engineering 30, no. 7 (2018): 1366-1385.

\bibitem{certisafe}
Sharma, Smita. "Certisafe, a novel Credential Authentication Process and System (CAPS)." U.S. Patent Application 14/349,363, filed March 12, 2015.

\bibitem{uniquecodes}
Malkawi, Mohammad Isam. "Counterfeit Prevention and Detection of University and Academic Institutions Documents Using Unique Codes." U.S. Patent Application 15/594,615, filed December 7, 2017.

\bibitem{parchment}
Digital Credential Service | Parchment", Parchment, 2018. [Online]. Available: http://www.parchment.com/. [Accessed: 04- Dec- 2018].

\bibitem{badgr}
"Badgr", Badgr.com, 2018. [Online]. Available: https://badgr.com/. [Accessed: 04- Dec- 2018].

\bibitem{mozillaopenbadges}
"Open Badges Homepage", Openbadges.org, 2018. [Online]. Available: https://openbadges.org/. [Accessed: 04- Dec- 2018].

\bibitem{appii}
"World's first blockchain career verification platform | APPII", APPII, 2018. [Online]. Available: https://appii.io/. [Accessed: 05- Dec- 2018].

\bibitem{gradba}
T. team, "Gradbase - Instantly Verify Qualifications", Gradba.se, 2018. [Online]. Available: https://www.gradba.se/en/. [Accessed: 17- Jan- 2018].

\bibitem{verify}
Zabar, Ed Adi. "Verification System." U.S. Patent Application 14/231,852, filed October 9, 2014.

\bibitem{accredible}
"Home", Accredible, 2018. [Online]. Available: https://www.accredible.com/. [Accessed: 05- Dec- 2018].

\bibitem{truerec}
Tummuru, Nethaji, Surbhi Sheth-Shah, Michael Kunzmann, Sanjay Shirole, and Jun Meng. "Decentralized credentials verification network." U.S. Patent Application 15/385,479, filed March 22, 2018.

\bibitem{bcdiploma}
BCDiploma", Bcdiploma.com, 2018. [Online]. Available: https://www.bcdiploma.com/. [Accessed: 05- Dec- 2018].

\bibitem{malaysia2018}
A. Abas, "University consortium set up to authenticate degrees using blockchain technology", Nst.com.my, 2018. [Online]. Available: https://www.nst.com.my/news/nation/2018/11/429615/university-consortium-set-authenticate-degrees-using-blockchain. [Accessed: 27- Jun- 2019].

\bibitem{indiachain}
Bansia, M., Murali, A., Murali, A. and Sen, S. (2019). India readies its biggest deep tech bet yet: a UPI-like blockchain platform | FactorDaily. [online] FactorDaily. Available at: https://factordaily.com/india-readies-upi-like-blockchain-platform/ [Accessed 27 May 2019].

\bibitem{nicosia}
"Academic Certificates on the Blockchain", UNIC Blockchain Initiative, 2018. [Online]. Available: https://digitalcurrency.unic.ac.cy/free-introductory-mooc/self-verifiable-certificates-on-the-bitcoin-blockchain/academic-certificates-on-the-blockchain/. [Accessed: 10- Jan- 2018].

\bibitem{hypercert}
Kim Hamilton Duffy, Learning Machine, Hypercerts: Blockcerts Revocation Improvements- By João Santos, Instituto Superior Técnico, 

\bibitem{echo}
Chen, S.X. and Team, E., Blockchain Based Professional Networking and Recruiting Platform.

\bibitem{uzhbc}
Gresch, Jerinas, Bruno Rodrigues, Eder Scheid, Salil S. Kanhere, and Burkhard Stiller. "The Proposal of a Blockchain-based Architecture for Transparent Certificate Handling."

\bibitem{channels}
Androulaki, Elli, Christian Cachin, Angelo De Caro, and Eleftherios Kokoris-Kogias. "Channels: Horizontal scaling and confidentiality on permissioned blockchains." In European Symposium on Research in Computer Security, pp. 111-131. Springer, Cham, 2018.

\bibitem{omniledger}
Kokoris-Kogias, Eleftherios, Philipp Jovanovic, Linus Gasser, Nicolas Gailly, Ewa Syta, and Bryan Ford. "Omniledger: A secure, scale-out, decentralized ledger via sharding." In 2018 IEEE Symposium on Security and Privacy (SP), pp. 583-598. IEEE, 2018.

\bibitem{plasma}
Joseph Poon and Vitalik Buterin. "Plasma: Scalable Autonomous Smart Contracts" Available at: [https://plasma.io/]

\bibitem{rootstock}
RSK - Smart Contract Platform Secured by the Bitcoin Network ", Rsk.co, 2019. [Online]. Available: https://www.rsk.co/. [Accessed: 30- Jun- 2019].

\bibitem{elements}
"Elements", elementsproject.org, 2019. [Online]. Available: https://elementsproject.org/. [Accessed: 30- Jun- 2019].

\bibitem{crosschain}
Deng, Liping, Huan Chen, Jing Zeng, and Liang-Jie Zhang. "Research on Cross-Chain Technology Based on Sidechain and Hash-Locking." In International Conference on Edge Computing, pp. 144-151. Springer, Cham, 2018.

\bibitem{multiplebc}
Kan, Luo, Yu Wei, Amjad Hafiz Muhammad, Wang Siyuan, Gao Linchao, and Hu Kai. "A multiple blockchains architecture on inter-blockchain communication." In 2018 IEEE International Conference on Software Quality, Reliability and Security Companion (QRS-C), pp. 139-145. IEEE, 2018.






















\bibitem{bittertobetter}
Barber, Simon, Xavier Boyen, Elaine Shi, and Ersin Uzun. "Bitter to better—how to make bitcoin a better currency." In International Conference on Financial Cryptography and Data Security, pp. 399-414. Springer, Berlin, Heidelberg, 2012.

\bibitem{hbr}
ansiti, M. and R. Lakhani, K. (2017). The Truth About Blockchain. [online] Harvard Business Review. Available at: https://hbr.org/2017/01/the-truth-about-blockchain [Accessed 1 Jul. 2019].


\end{thebibliography}
\end{document}